\documentclass[10pt, a4paper]{article}

\usepackage{authblk}
\usepackage{graphicx}
\usepackage{listings}
\usepackage{blindtext}
\usepackage{enumitem}
\usepackage{backnaur}

\title {Clarification of Video Retrieval Query Results by the Automated Insertion of Supporting Shots}

\author {Sean Butler}

\affil{Department of Computer Science and Creative Technologies, UWE Bristol, BS16 1QY, UK}

\date{}

\begin{document}

\maketitle

\begin{abstract}
Computational Video Editing Systems' output video generally follows a particular form, e.g. conversation or music videos, in this way they are domain specific. We describe a recent development in our video annotation and segmentation system to support general computational video editing in which we derive a single generic editing strategy from general cinema narrative principles instead of using a hierarchical film grammar. We demonstrate how this single principle coupled with a database of scripts derived from annotated videos leverages the existing video editing knowledge encoded within the editing of those sequences in a flexible and generic manner. We discuss the cinema theory foundations for this generic editing strategy, review the algorithms used to effect it, and go on by means of examples to show its appropriateness in an automated system.


\end{abstract}



\section {Introduction}
Researchers have applied Computational Video Editing (CVE) across a variety of domains, including news programmes \cite{Low1997AutomaticNewsVideoParsing}, sports \cite{DELAKIS2008142}, show trailers \cite{Sack1994IDICAV}, or interviews \cite{Leake:2017:CVE:3072959.3073653}. Systems generally make use of a distinctive, limited set of editing principles to assemble their outputs for that domain. While textbooks, theorists and practitioners \cite{arijon1991grammar}  \cite{Katz1991Shot} discuss complex film grammars applied in the general case, automatically targeting this general case is properly hard due to an explosion in permutations and we are unfamiliar with computational intelligence applicable in the general case.

Grammatical film editing techniques such as the master shot, which can result in a movie where the structure is visible to the audience, have fallen out of fashion. In narrative driven movies this is generally unacceptable for the audience to see the structure of the editing, except when it isn't, see \cite{Anderson} \cite{Greenaway1980Falls} which both draw attention to their formal structure as part of the storytelling. Our aim has been to find generic automatic editing strategies which are both applicable across multiple domains and which are not apparent to the audience each time they are used with the objective of developing systems which use a naturalistic editing style.

Advances in computational intelligence and natural language processing \cite{NLPSurvey2014} mean systems can now translate natural language into machine readable formats such as the formal and logical Conceptual Graphs \cite{Sowa1984ConceptualSI} and systems can regard images and create accurate descriptions of their content \cite{Kadir2001}.

We have created a video retrieval and editing system which uses human readable queries to assemble video sequences from a database of footage. The database of video is processed and annotations are combined and segmented so that they describe the visual content of the frame over intervals. A single global editing strategy derived from narrative and film theory, based on cause and effect using the preconditions and postconditions of a situation is used to recombine video in response to external queries.

The primary benefits of this approach are:
\begin{itemize}
\item We can extract and apply the existing editing knowledge built into reused/re-edited video sequences.
\item The video produced follows readable editing patterns without visible structure
\end{itemize}

Also, the system is easily interfacable with other Computational Intelligence technologies, and the sequence processing being derived from set and graph theory contribute to a formal understanding of film theory and a definition of the unit of film grammar.

\section{Background}

\subsection{Context and Meaning in Film}
\begin{figure}
	\centering
	\includegraphics
			[width=0.5\textwidth]{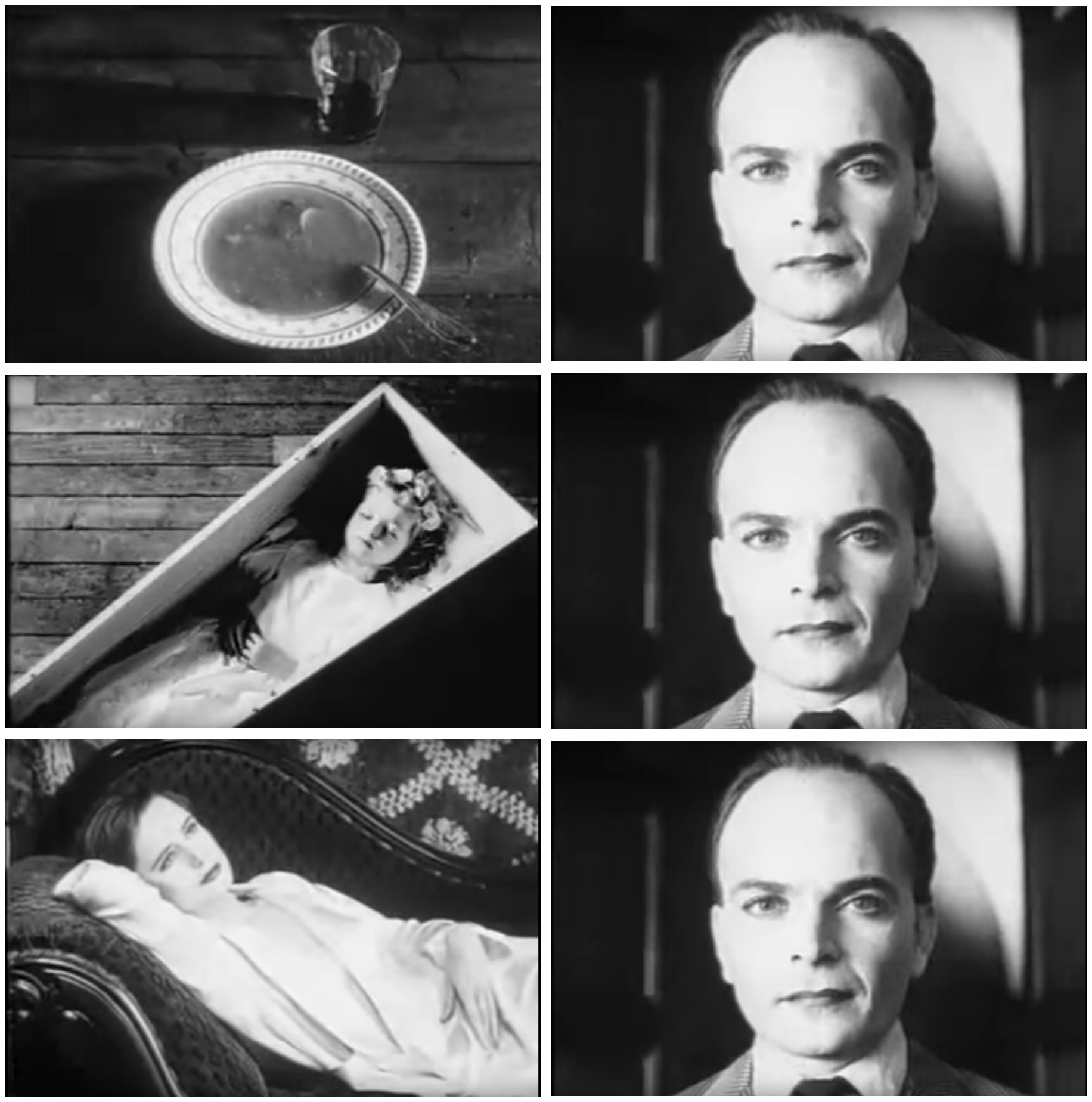}
	\caption{Stills from a demonstration of The Kuleshov Effect, where the perceived emotion changes depending on the context}
  \label{fig:Kuleshov}
\end{figure}

In the 1920s Kuleshov showed that the effect of perceiving a continuous reality when shown a rapid sequence of frames is so effective and so closely mimics the conscious experience that it also works across sequences of images that don't have similar contours. This has since become known as ``The Kuleshov Effect'' \cite{Mobbs2006TheKE}. In Kuleshov's experiment, audiences perceive hunger, grief and attraction in the same shot of an actor’s face that has been juxtaposed with different successive shots. As can be seen in Figure \ref{fig:Kuleshov}, \cite{KuleshovYoutube} the effect is revealed to be an illusion but when presented with the sequences in isolation audiences perceive the emotion.

The Kuleshov Effect may be explained by saying the audience uses their imagination and experience to create a fiction which accounts for the images before them. The audience reads the film assuming the shots are related and discovers or invents \cite{kawin1992movies} a reason or for that relation.
Cognitive scientists \cite{Magliano2011TheIO} have studied the phenomena. Explanations of the effect have ranged widely over, for example, theories of Point of View, Consciousness and Eye Movements. Film students are commonly advised to consider the camera as an additional person in the scene observing the other protagonists and edit according to an observer's behaviour.

\subsection{Meaning, Retrieval and Editing}

Due to the Kuleshov Effect we already know that should the context of a shot change, then its meaning will also likely change.  All cases of retrieval, except the first solitary one of a shot retrieved into isolation also change the context. Therefore retrieval also changes the meaning. After retrieval any editing operations which change the context will also alter the meaning.

The permutations of meaning change are many and would  present a significant generation and indexing challenge. It follows that a system designed to perform automatic re-editing of video footage will run into significant problems should it maintain data about the meaning of shots or sequences. We must therefore focus on managing the content of the shots rather than the meaning of those shots.

\subsection{Automatic Video Structure Detection}
Since \cite{Zhang1993AutomaticPO} many authors \cite{Lienhart1999ComparisonOA} have developed algorithms to determine the structure of video. These approaches generally use a mathematical method to summarise the difference between neighboring frames or those in short succession. In doing so they detect the linear structure of a shot sequence. We envisage strategies such as those we are developing to be integrated with approaches to create hybrid systems.

\subsection{The Evolution of Annotation In Computational Video Editing}
Operating on video containing complex engineering activities Parkes \cite{PARKES1989171} showed the effectiveness of Conceptual Structures (CGs) \cite{Sowa1984ConceptualSI} as an notation to describe the content of an instructional video.  CGs support description and also processing for transitions suitable for the reediting of instructional videos for different learners.

\begin{quote}
``Conceptual graphs (CGs) are a system of logic based on the existential graphs of Charles Sanders Peirce and the semantic networks of artificial intelligence. They express meaning in a form that is logically precise, humanly readable, and computationally tractable.'' \cite{SowaCGSite2005}
\end{quote}

The Stratification System \cite{Smith1992TheSS} (an early attempt to maintain and visualise layered descriptions of video content) stores keywords over time and is focussed on the interface so a user can manipulate the descriptions easily for video annotation. Butler \& Parkes \cite{Butler1996FilmicSD} presented a video annotation and editing system which also used keywords over time and provided a visual interface, in this case the user could join shots together to create a space-time diagram and see the structure of the edits.

Sack and Davies \cite{Sack1994IDICAV} describe an annotation and repurposing of Star Trek episodes by means of story plans and content annotations to create simplified or short sequences following a particular dramatic structure. Davies \cite{Davis1994KnowledgeRF} considers a semantic approach that handles the issue of context and meaning by arguing for a notation that can encompass the meaning in isolation and the meaning in context.

Butler and Parkes \cite{ButlerParkes1997FilmSequenceGenerationStrategies} described a system which also uses CGs to describe video content coupled with layers of intervals \cite{Allen1983MaintainingKA}. The main development was the use of the generalisation of Conceptual Graphs by means of an ontology along with a range of idiomatic editing rules such as the 180 degree rule, parallel actions and subjective shots.

Nack \cite{NackParkes1997AutomatedEditingTheme} described theme oriented video editing using planning form and theme strategies to achieve comedic effect. Focussing on argued opinion documentaries, Bocconi and others \cite{1521610} go on to show how higher level annotations and recombination schemas can be used to generate specific forms of video production.

Various algorithms were investigated \cite{Shipman2003GIMMLVS} for the creation of multi-level video summaries in the home movies domain, using a hybrid of technologies including mathematical shot boundary detection, external annotation and defined structures of summaries.

More recent computational approaches use a variety of technologies including machine learning and computer vision. For example \cite{Leake:2017:CVE:3072959.3073653} with a focus on cross-cutting between complete captures of scenes from multiple angles for multi party conversation. Merabiti \cite{Merabti2016VirtualDirectorHMM} created a hybrid system making use of Hidden Markov Models to learn editing from source videos which have been annotated by a person.

In summary, many researchers working at the current state of the art in computational video editing make use of a hybrid approach. That is of machine vision and machine learning along with symbolic annotations. In some cases the sytems generate annotations and in others they use external annotations with the system operating on the annotations. As you can see systems have evolved from the purely symbolic logic programming approaches in the 90s through the application of computer vision technologies to the modern hybrid approachs of machine learning and symbolic annotation combined.

\section{Deriving A Generic Video Editing Strategy}
A short review of some major narrative principles follows. By focussing on editing and narrative principles which are applicable in any/all contexts and are unrelated to specific events or portrayals we have formulated a generic editing principle.


Deus Ex Machina or God from the Machine refers to the unsatisfying resolution of a story which otherwise would have been impossible for the protagonists to resolve that way. Historically the gods might have intervened in the story and entered from above the stage by means of a mechanism \cite{AristotlePoeticsTrans}. In modern stories it refers to sudden or late powerful interference, impossible deductions, coincidence, dreams or other similar unsatisfying resolutions. For automatic video this means we cannot show an event which relies on an event, object or other entity within the narrative that has not appeared prior to this point.

One of the expressions of ``the rule of three'' in Hollywood movies is that you show them three times: First show the audience that you are going to do it, then you show it happening, then you show them that you did it. Historically this may have been taken quite literally. More recently we don't show all three phases of an event. In practice the speed of the shot and the complexity of the event have an impact on whether we show 1 or 2 or 3 of the parts.

Consider the sequence in ``Shaun of the Dead'' where the protagonist and his buddy are discussing how they are going to survive the zombie apocalypse. Several refinements of their plan are discussed and the sequences are shortened each time. The first iteration of the plan is shown very quickly and understandable because each event is shown with multiple shots. Those shots being the preconditions and/or postconditions of an event in the sequence. Examine the car sequence within the first iteration of the plan and we can see how taking the car is shown in great detail with several successive shots of the preconditions.

\begin{figure*}
\centering
\includegraphics
		[width=\textwidth]
		{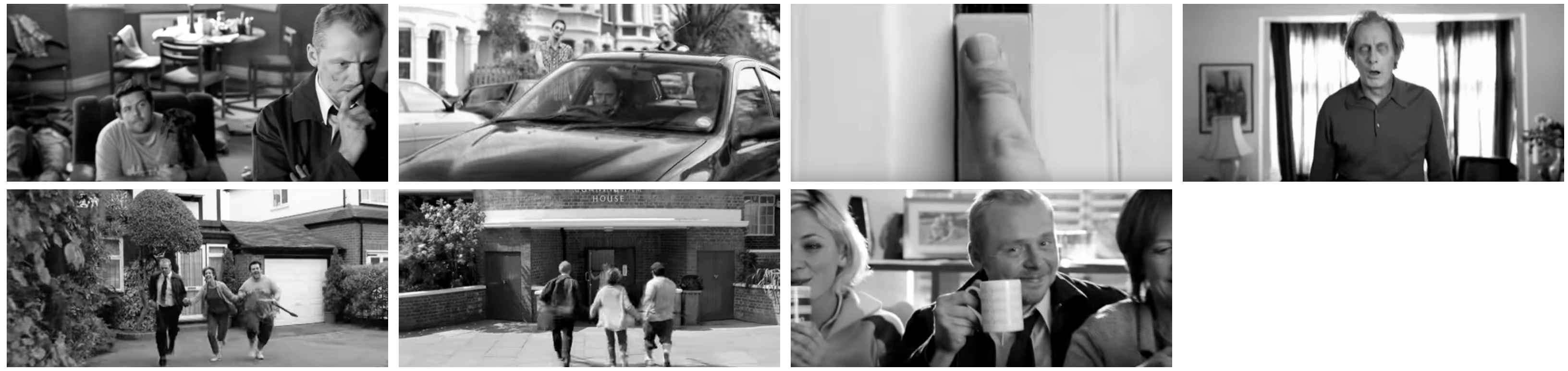}
	\caption{Stills from the major shots of the ``Shaun of the Dead'' Survival Planning}
	\label{fig:ShaunPlanning}
\end{figure*}

\begin{figure*}
\centering
\includegraphics
		[width=\textwidth]
		{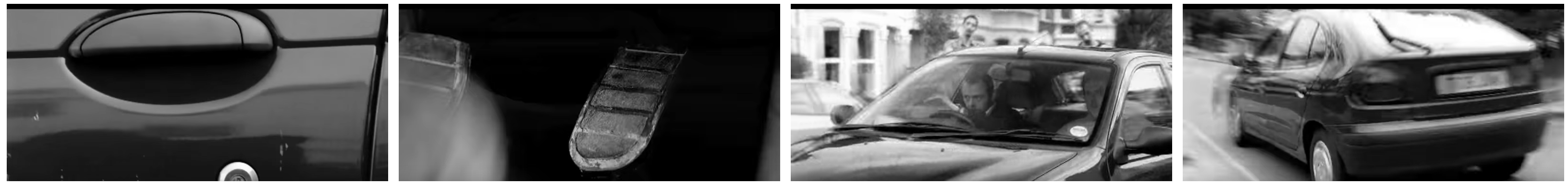}
	\caption{Stills from each shot of the ``Shaun of the Dead'' car subplan}
	\label{fig:CarSubplan}
\end{figure*}

\subsection{Structural Psychology of Cinema}

J.M.Carroll \cite{Carroll80PsycCinema} discusses film psychology and grammar in great detail. In his psychological and grammatical treatment of film Carroll separates the visualisation of actions in film into preparatory actions and focal actions also focal actions and event terminations. He then builds these as separate non-terminals into his partial film grammar.

{\small

\begin{bnf*}
\bnfprod{sequence}
{\bnfpn{action} \bnfor \bnfpn{action} \bnfsp \bnfpn{action} }\\
\bnfprod{action}
{ \bnfpn{preparatory action} \bnfsp \bnfpn{focus-action}}\\
\bnfprod{action}
{ \bnfpn{focus action} 	\bnfsp \bnfpn{event-termination}}\\
\end{bnf*}

\par}

In his transformational grammar, rules may be used to change a sequence of actions into a sequence of shots depending on the interrelationships between the shots and the desires of the editor/director. Throughout the work Carroll presents examples of additional rules for specific scenarios, such as master scene, parallel actions and subjective shot. Being a partial system, and due to the recursive rules, what is not clear is how the rules can be motivated to achieve a specific editing goal.

\subsection{A Theory of Generic Video Editing}

Each of the above strategies is a story telling technique, and can therfore be used as a general principle to be applied repeatedly across an entire video.

\begin{itemize}

\item We should omit events that are unnecessary for the audience to understand the story.

\item Audiences fill in the gaps in missing, truncated or summarised events.

\item If we cant show the event itself, then we can show the preconditions and/or the postconditions and the audience will conclude the event has occured.

\item If we wish to show an event more clearly, then we can show the preconditions and/or the postconditions bracketing the event itself.

\end{itemize}

Assuming that video editing is an interleaving of events to create a linear sequence, to show an event in film we have a variety of choices:

\paragraph{Preconditions the Situation then the Postconditions} This is the clearest form and is the least challenging for an audience to understand. It takes longer, but this approach would be appropriate when the event is complicated, fast, or not obvious to an audience.

\paragraph{Preconditions then the Situation:}This works for most circumstances and is appropriate where the results of the event are well understood because the audience is familiar with the domain.

\paragraph{Situation then the Postconditions:}This works for most circumstances and is appropriate where the result of the event is important for the later content.

\paragraph{Preconditions then the Postconditions:}Show a plane taking off, then someone walking into an office. We show someone putting a posting a letter then show another person reading a letter. We don't need to show it being written, being received, or travelling between the places.

Each of these approaches has similarities and differences in their interpretation by an audience, which is a topic for future work, but all serve to communicate more clearly.

\subsection {Production Rules}
The effect of the principles dissussed can be more succinctly described by the following production rules. Each terminal and non-terminal symbol can ultimately be replaced by a situation from the database.

{\small

\begin{bnf*}
\bnfprod{action}
{\bnfpn{precondition} \bnfsp \bnfpn{action} \bnfsp \bnfpn{postcondition} }\\
\bnfprod{action}
{ \bnfpn{precondition} \bnfsp \bnfpn{action}}\\
\bnfprod{action}
{ \bnfpn{action} \bnfsp \bnfpn{postcondition}}\\
\bnfprod{action}
{\bnfpn{precondition} \bnfsp \bnfpn{postcondition} }\\
\end{bnf*}

\par}

These are not shots. The footage showing the action desired becomes shots only after it has been extracted from a longer sequence (of situations) and combined into the resulting edited sequence.


\section{Implementation}

\subsection{Overview}

\begin{figure}
   	\centering
   	\includegraphics [width=0.7\columnwidth]{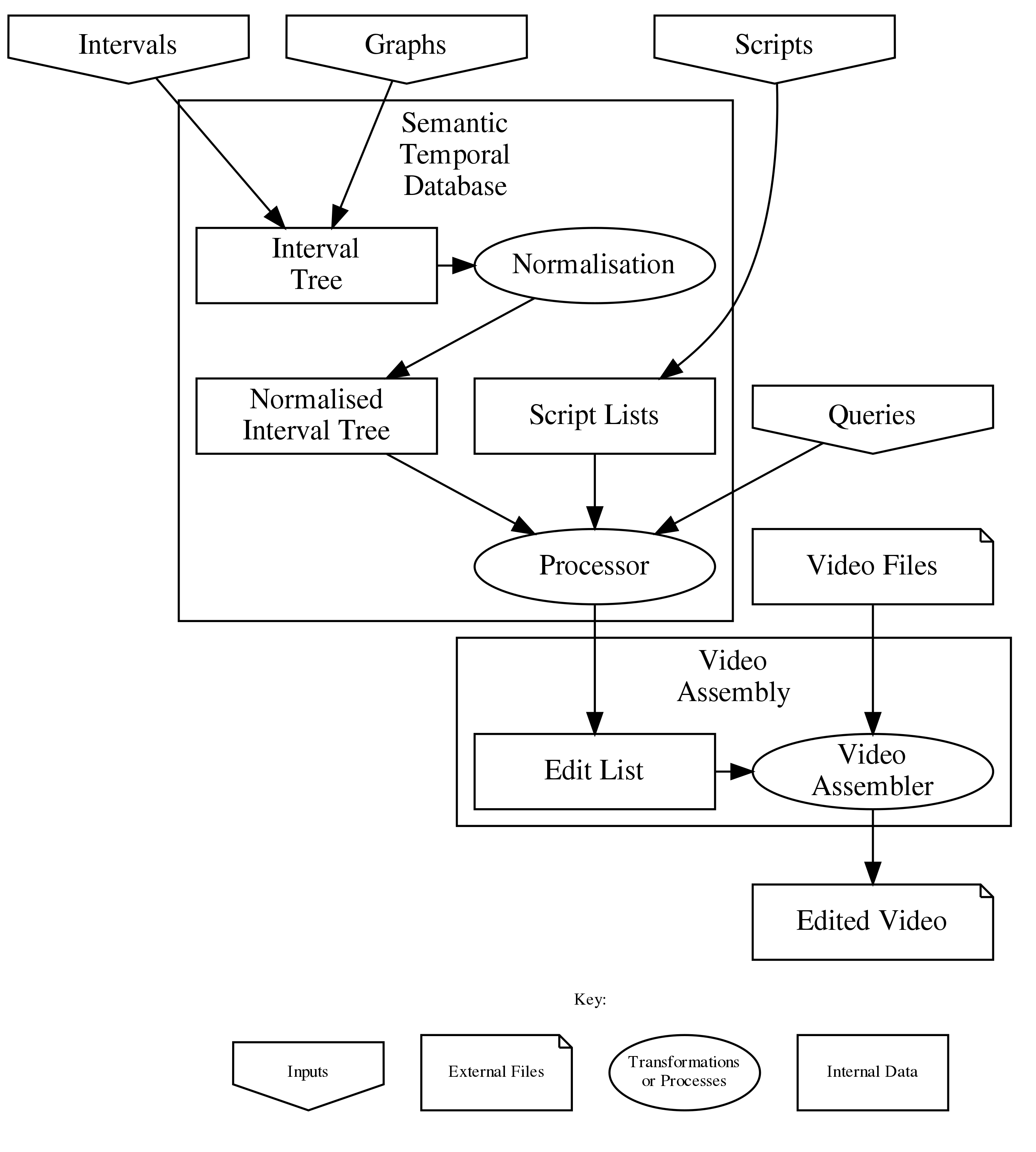}
   	\caption{Overview of System Dataflows}
   	\label{fig:SystemOverview}
\end{figure}

In the system overview (Figure  \ref{fig:SystemOverview}) you can see each of the major data and processes which occur in the system. `Processor' is where the sequences are generated from the interval tree scripts in the database when presented with a query. The final assembly is carried out in a separate module from a list of cuts, similar in principle to a non linear editor.

The system maintains a set of CGs. Each CG represents a description of the visible content of some contiguous frames of video and is accompanied by filename and time data which represents the interval for which it is said to hold.

Multiple overlapping intervals with their associated CGs are inserted in an interval tree which affords time efficient query operations.

During an insert operation, if no overlap exists, then the new interval and CG annotation is inserted. If an overlap is found, then new intervals are generated from the intersection of those overlapping such that the overlap is removed. Corresponding to each new interval, a union of each CG associated is created to form a new concept describing the content of the image during that interval.

A script consists of an ordered sequence of CGs and is used to indicate that some events are preconditions and some are postconditions of others. The exact timings are not important rather their order is, in this way the scripts resemble a traditional AI plan or script. \cite{Schank1975ScriptsPA}

As its impossible for a  system to know all the possibilities for meaning change as context changes, we are not implementing rules to choose intervals based on their conceptual description and its relation to the event. Though other systems have been implemented which operate in this way \cite{NackParkes1997AutomatedEditingTheme}. Where possible the system chooses from the intervals in the database which match those in the scripts. Where more than one possibility exists we choose one with the closest relationship to the event.

Where the retrieval action results in a choice of more than one precondition or more than one post condition for a given event built the system to choose the interval which is closest in sequence to the event. Similar scripts can be generated for any linear sequence of situations. Watering the Plants, Making Coffee, Visiting Friends, Assembling an Assault Rifle, Operating a Micrometer, etc

\section{Results}

Looking at a specific examples we can show how the preconditions post conditions event rules can be used to generate sequences which clearer and  more understandable.

Each concept in the query is tested against the conceptual database which describes the video content available and if the database contains the concepts requested, they are retrieved. In the examples here we have forced the system to retrieve results with and without applying the editing principles for comparison.

Consider the following example querie in which we ask the system to retrieve a short video of a person visiting.

\begin{verbatim}
[visit]<-(AGNT)<-[person]
\end{verbatim}

\begin{figure}
\centering
\includegraphics
		[width=0.25\textwidth]
	{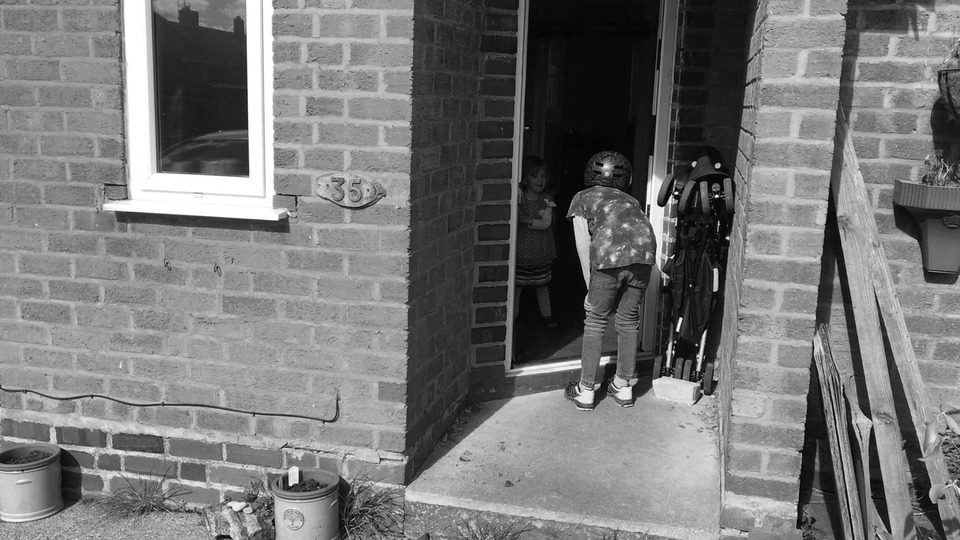}
\caption{Still from the shot resulting from ``Visit'' Query}
\label{fig:VisitWithout}
\end{figure}

\begin{figure}
\centering
\includegraphics
		[width=0.5\textwidth]
	{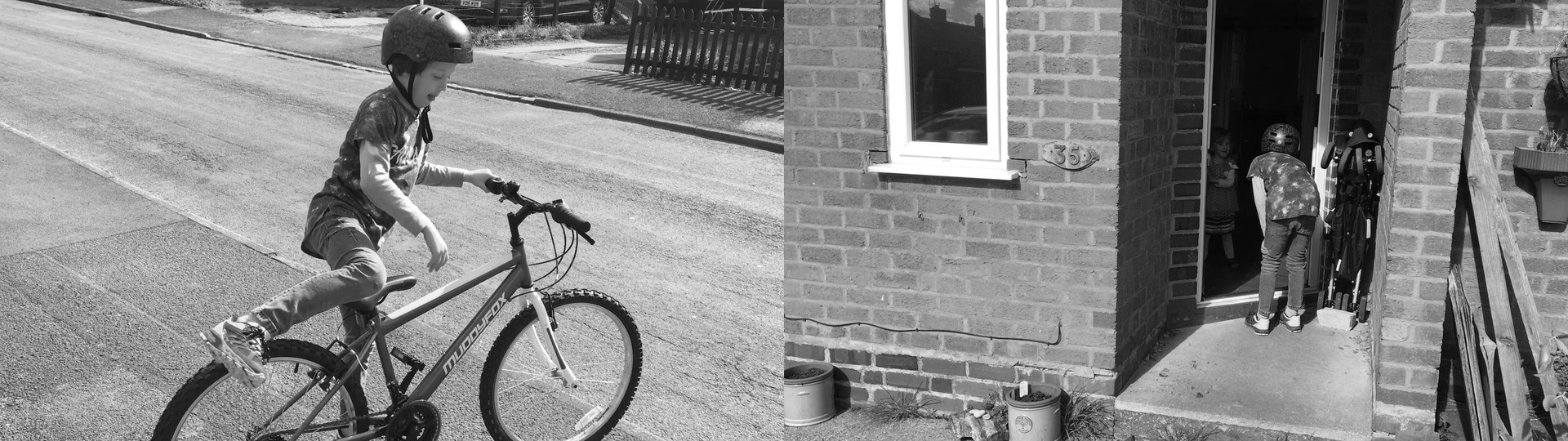}
\caption{Stills from each shot resulting from ``Visit'' Query, where we see the insertion of a preconditions shot}
\label{fig:VisitWith}
\end{figure}

You can see the results in Figure \ref{fig:VisitWithout}. The system finds the visit script and retrieves a shot. In Figure \ref{fig:VisitWith} it looks for a precondition and inserts that into the generated edit list.

\begin{verbatim}
[visit]<-(AGNT)<-[person]
[drink]<-(AGNT)<-[person]
\end{verbatim}

You can see the results in Figure \ref{fig:VisitWithout}. The system finds the visit script and the cup of coffee script.

In Figure \ref{fig:VisitWith} the system looks for a precondition and inserts that into the generated edit list.

\begin{figure}
\centering
\includegraphics
[width=0.5\textwidth]
	{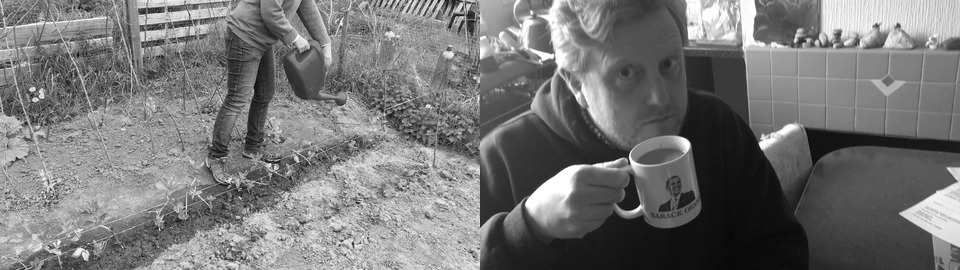}
\caption{Stills from each shot of the results from ``Allotment and Drink'' Query.}
\label{fig:AllotmentCuppaWithout}
\end{figure}

You can see the results of the following query in Figure \ref{fig:AllotmentCuppaWithout}.

\begin{verbatim}
[irrigate]-
    <-(AGNT)<-[person]
    <-(OBJ)<-[plants]
[drink]<-(AGNT)<-[person]
\end{verbatim}

\begin{figure*}
\centering
\includegraphics
[width=\textwidth]
	{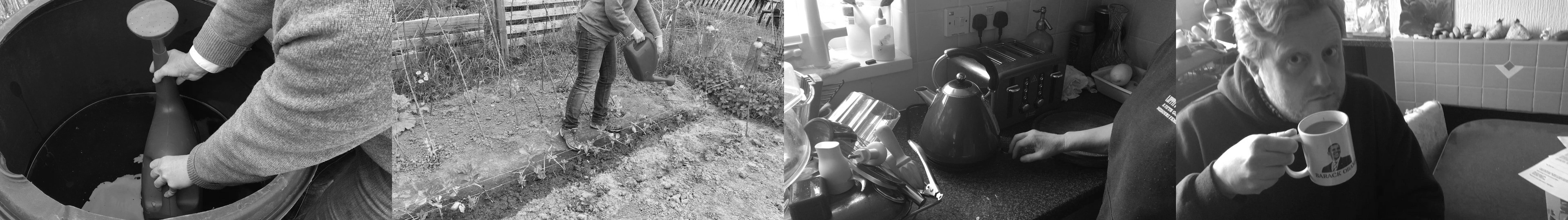}
\caption{Stills from each shot of the results from ``Allotment and Drink'' Query, where we see the insertion of a preconditions shot for the allotment watering, and a preconditions shot for the drink.}
\label{fig:AllotmentCuppaWith}
\end{figure*}

Again with the pre/post conditions editing rule applied in Figure \ref{fig:AllotmentCuppaWith}.

\begin{figure*}
\centering
\includegraphics
[width=0.5\textwidth]
	{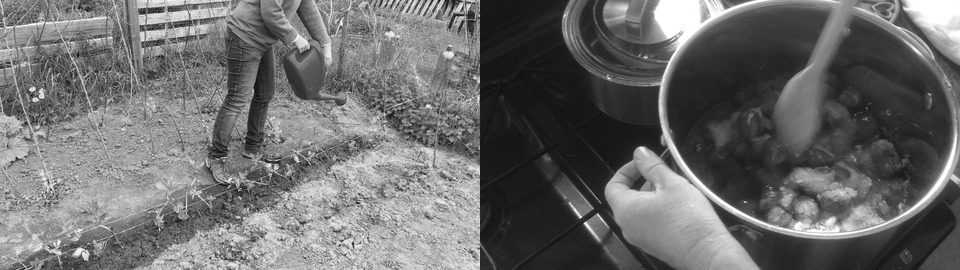}
\caption{Stills from each shot of the results from the ``Allotment and Canning'' Query.}
\label{fig:Query3Without}
\end{figure*}

You can see the results of the following query in Figure \ref{fig:Query3Without}.

\begin{verbatim}
[irrigate]-
    <-(AGNT)<-[person]
    <-(OBJ)<-[plants]
[canning]-
    <-(AGNT)<-[person]
    <-(OBJ)<-[fruit]
\end{verbatim}

\begin{figure*}
\includegraphics
[width=\textwidth]
	{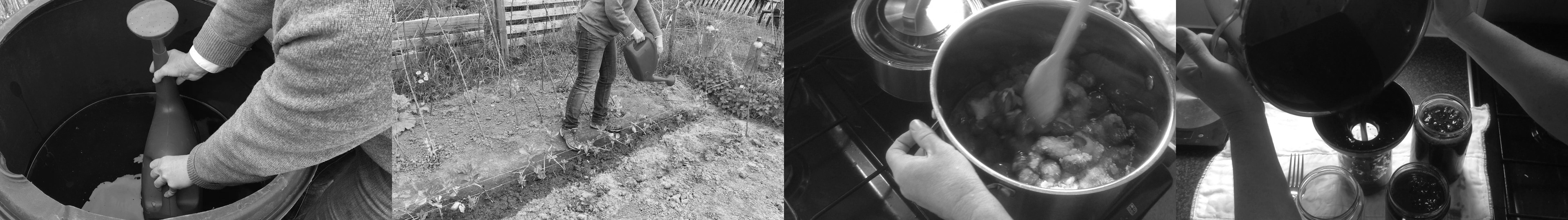}
\caption{Stills from each shot of the results from ``Allotment and Canning'' Query, with the preconditions and postconditions rules applied.}
\label{fig:Query3With}
\end{figure*}

Again with the pre/post conditions editing rule applied in Figure \ref{fig:Query3With}.

\section{Conclusion and Discussion}

The system assembles video sequences which match the concepts requested. The sequences assembled by the application of this single strategy are clearly more easily comprehensible than those without.

Scripts of events from normal/common causal chains often include events which though are preconditions for this particular instance of an event, aren't a precondition for the generic form. Additional annotations would fix this in the short term we aim to develop a system which doesn't need additional data beyond the image contents. The ordering of events is determined by the scripts which are derived from the content of human made video sequences. It is in this way we capture some of the `grammar' of film without having to construct specific rules. Applying the editing strategies outlined earlier, the system chooses a context of preconditions or postconditions for the events supplied. The combination of these additional shots with the desired event renders the overall sequence understandable.

Rather than the application of a hierarchical rule based system automated video editing and retrieval should focus generic methods for the assembly of context of shots.

The location of potential cuts in the database for the system to use is derived from the processing of overlapping intervals each associated with a concept which can be said to be true and holds overall the images in the interval. Each time the system chooses potential cut it does so to satisfy a goal on the sequence level, the places it can choose from are not limited to existing cuts in the sequence. The videos created are clear and understandable (though the rhythm/pacing is inconsistent). The unit of film grammar should not be considered as a shot bounded by cuts, but instead the `interval'. Where an interval is sequence of frames with continuous contours and constant visible content. There is also an argument for the unit being the sign-interval within the frame, but further work is needed, and that if this is the case, then the interval becomes a non-terminal within a more detailed system.

\section {Future Work}

Our immediate research goals lie in two areas, machine learning and further analysis of overlapping intervals:

To apply the editing rules to a much larger body of video and scripts, we plan to extend our system into a larger toolchain. To interface the editing system with a source of automatically generated annotations of video image content along with a feedback user interface so an operator can indicate the quality of the results generated.

To automatically derive the scripts from existing sequences of video and to automatically derive a hierarchy from the multiple overlapping intervals of concepts describing the objective content of the video. This hierarchy would be used to more flexible support the application of the singular editing rule in different scenarios.

This study has focussed on investigating computational approaches to generic video editing using concept boundaries as potential cut opportunities. Cutting at concept boundaries is however, only one approach, another is to cut in the middle of an action, to this end we plan to investigate ``The Invisible Cut'' technique where shot boundaries occur mid action as an event sequences from one internal phase to another.

\bibliographystyle{unsrt}

\bibliography{Article}


\end{document}